\documentclass{pasj00}
\draft

\begin{document}
\SetRunningHead{Author(s) in page-head}{Running Head}
\Received{2000/12/31}
\Accepted{2001/01/01}



%


\title{ Particle Propagation in the Galactic Center and
Spatial Distribution of Non-Thermal X-rays}

\author{Vladimir \textsc{Dogiel}$^{1,2}$, Dmitrii {\sc Chernyshov}$^{2,3}$,
Takayuki {\sc Yuasa}$^4$, Kwong-Sang \textsc{Cheng}$^5$,   Aya
{\sc Bamba}$^{1}$, Hajime {\sc Inoue}$^{1}$, Chung-Ming
 {\sc Ko}$^6$, Motohide {\sc Kokubun}$^{1}$, Yoshitomo  {\sc
Maeda}$^{1}$, Kazuhisa {\sc Mitsuda}$^{1}$, Kazuhiro {\sc
Nakazawa}$^4$, and Noriko Y. {\sc Yamasaki}$^1$}
 \affil{$^1$Institute of Space and Astronautical
Science, 3-1-1, Yoshinodai, Sagamihara, Kanagawa, 229-8510, Japan}
  \affil{$^2$P.N.Lebedev Institute, Leninskii
pr, 53, 119991 Moscow, Russia, dogiel@lpi.ru}
\affil{$^3$Moscow Institute of Physics and Technology, Institutskii lane, 141700 
Moscow Region, Dolgoprudnii, Russia} \affil{$^4$Department of
Physics, School of Science, The University of Tokyo, 7-3-1 Hongo,
Bunkyo-ku, Tokyo 113-0033}
  \affil{$^5$Department of Physics,
University of Hong Kong, Pokfulam Road, Hong Kong, China}
\affil{$^6$Institute of Astronomy, National Central University,
Jhongli 320, Taiwan}

\maketitle
\begin{abstract}
We showed that if the non-thermal emission from the Galactic center
in the range $14-40$~keV is due to inverse bremsstrahlung
 emission of subrelativistic protons,  their interactions with
 hot and cold fractions of the interstellar medium are equally
 important. Our estimation show that about 30\% of the total
 non-thermal flux from the GC in the range $14-40$~keV is generated in
 regions of cold gas while the rest is produced by proton
 interaction with hot plasma. From the spatial distribution of 6.7
 keV iron line we concluded the spatial distribution of hot plasma is
 strongly non-uniform that should be taken into account in analysis of protons
propagation in the GC. From the  Suzaku data we got independent
estimates for the diffusion coefficient of subrelativistic protons
in the GC, which was in the range $ 10^{26} - 10^{27}$
cm$^2$s$^{-1}$
\end{abstract}


\section{Introduction}

This is a final paper of the series (\cite{cheng1,cheng2,dog08};
Dogiel et al. 2009abc) on energetic processes in the Galactic
center (GC). In these papers we presented our interpretation of
X-ray and gamma-ray  emission from the Galactic center. We
supposed that all these phenomena had common origin and were
initiated by accretion processes onto the central supermassive
black hole.

We showed that the the observed X-ray continuum and line emission
from the GC might be produced by a flux of subrelativistic protons
which resulted from an unbounded part of stars accreted onto the
central black hole. We estimated the average energy of escaping
protons to be about 100 MeV in order to produce a flux of hard
X-ray emission in the energy range $14-40$~keV as observed by
Suzaku/HXD \citep{yuasa}.

The quasi-stationary production rate of subrelativistic protons
hardly exceeds $Q\sim 2\times 10^{45}$ protons s$^{-1}$ for the
frequency of star capture $\sim 10^{-5}-10^{-4}$ years$^{-1}$
\citep{syer,don} and a fraction of escaped matter equaled $\sim
50$\% of star masses \citep{ayal}.

The unbounded fraction of stars escapes with subrelativistic
velocities which correspond energies about  100 MeV for protons
and 50 keV for electrons (see \cite{dog09}).  The numbers of
protons and electrons equal to each other. Therefore, fluxes of
bremsstrahlung hard X-ray emission produced by protons and by
electrons equal to each other too. But the lifetime of electrons
with energies 50 keV is in about five orders of magnitude smaller
than that of 100 MeV protons and the electron bremsstrahlung flux
is significant for a  short time just after a capture event. Since
it is supposed that  relatively long time has passed  after the
last star capture, electron bremsstrahlung radiation is
negligible.

Our goal is to estimate the  spatial diffusion coefficient, $D$,
of subrelativistic particles near the GC, whose value is unknown.
Below we derive its value from the spatial distribution of hard
X-ray emission near the GC as observed by {Suzaku}.

In section 2, we summarize an energetics of the Galactic Center
diffuse emission. In section 3, we carry forward our model using
non-uniform target gas distribution. Section 4 is devoted to
explain how the model and the HXD data can be compared. In section
5 and 6, spatial distribution of 6.7~keV Fe line emission and hard
X-ray continuum are compared with ones expected from the present
model, deriving a confinement on a diffusion coefficient for the
sub-relativistic protons.

 \section{Components of Hard X-Ray Emission from the GC in the range $14-40$~keV}

\citet{koya2}  reported that the hard X-ray spectrum in the range
$2-10$~keV from the GC
 is naturally explained by a 6.5~keV-temperature plasma
 plus a power-law component with the photon index of
$\Gamma=1.4$. Latter on, \citet{yuasa} found a prominent hard
X-ray emission in the range from 14 to 40 keV whose spectrum is a
power law with the spectral index ranging from 1.8 to 2.5.

Different processes may contribute to the total emission from the
GC.

{\it Thermal emission}. The emissivity of thermal bremsstrahlung
of a hot plasma $\varepsilon$ can be
 estimated from the equation (see e.g. \cite{ginz})
 \begin{equation}
4\pi\varepsilon=1.57\times 10^{-27}n^2(r)\sqrt{T(r)}~\mbox{erg
cm$^{-3}$s$^{-1}$}
\end{equation}
where the temperature is in Kelvin degrees.

As follows from the analysis of { Suzaku} data the plasma
temperature in the GC is constant and equals 6.5 keV
\citep{koya2}. Then the thermal emissivity in the GC is
\begin{equation}
4\pi\varepsilon\simeq 1.27\times 10^{-23}n^2(r)~\mbox{erg
cm$^{-3}$s$^{-1}$}
\end{equation}

\citet{dog_pasj} presented results of calculation for  the
$0.55^\circ\times0.55^\circ$ central region and showed that for
the plasma density $n=0.2$ cm$^{-3}$ and the temperature 6.5 keV
the total thermal flux from this region in the energy rage $14-40$~keV
was about $F_{\rm th}\simeq 2\times 10^{36}$ erg s$^{-1}$. The
total non-thermal flux was estimated by the value $F_{\rm
nth}\simeq 3\times 10^{36}$ erg s$^{-1}$.

{\it Non-thermal component}. \citet{koya09} found that a
combination of the Fe XXV-K$\alpha$ and Fe I-K$\alpha$ fluxes  was
proportional to the non-thermal continuum flux  in the 5 - 10 keV
energy band. They concluded that the total non-thermal flux
consists of the two components, one of which is proportional to
the intensity of 6.7 keV line, and, therefore, is produced in
regions of hot plasma, while the intensity of the second component
is proportional to that of the 6.4 keV line and is generated in
regions of molecular gas. \citet{warw} reported a similar
conclusion obtained from the XMM-Newton observations.

In the model of X-ray production by subrelativistic protons this
relation is naturally explained (see Dogiel et al. 2009bc).
Continuum and line emission is produced in this case by
interactions of subrelativistic protons with the background plasma
and the molecular hydrogen in the GC.

 In spite of
relatively small radius ($r\sim 200$ pc) the inner Galactic region
contains about 10\% of the Galaxy's molecular mass, $\sim M_{\rm
H_2}\simeq(7-9)\times 10^7~M_\odot$.  Most of the molecular gas is
contained in very compact clouds (see the review of
\cite{mezger}). Since the total mass of the molecular gas is much
larger than that of the hot plasma, $M_{\rm pl}\sim 4\times
10^5~M_\odot$, one may assume that most of inverse bremsstrahlung
flux produced by protons would come from regions filled by the
molecular gas. However, as \citet{dog_p3} showed, the mean free
path of protons in dense molecular clouds is very short since the
diffusion coefficient inside molecular clouds is quite small.
Therefore, only a small fraction of the molecular hydrogen is
involved into processes of inverse bremsstrahlung radiation. Thus,
for the molecular cloud Sgr B2 with the mass about $10^6~M_\odot$
it follows from calculations of \citet{dog_p3} and the { Suzaku}
data (see \cite{koya3}) that the total bremsstrahlung flux in the
range $2-10$~keV is about $F_{\rm 2-10}\simeq  10^{35}$ erg
s$^{-1}$. The total flux of the 6.4 keV line emission from Sgr B2
is about $F_{\rm 6.4}\sim 1.2\times 10^{34}$  erg s$^{-1}$ that
gives the ratio $F_{\rm 2-10}/F_{\rm 6.4}\sim 8$. For higher
energies of X-ray photons our calculations show that the ratio of
$F_{\rm 14-40}/F_{\rm 6.4}\sim 6.4$.

The total flux of the 6.4 keV line from the GC can be estimated
from the ASCA data  \citep{maeda8} and for the region
$0.55^\circ\times 0.55^\circ$ it  is about $\sim 1.3\times 10^{35}$
erg s$^{-1}$, then we expect that the total non-thermal
bremsstrahlung flux from proton interaction with  the molecular
gas is about $10^{36}$ erg s$^{-1}$, and  70\% of the total
non-thermal bremsstrahlung emission in the range $14-40$~keV, $\sim
2\times 10^{36}$ erg s$^{-1}$, is generated in regions filled with
the hot plasma.

 \section{Qualitative effect of the plasma density variations}
 In this section we present a qualitative effect of  plasma
 density variations on the proton distribution in the GC.
 We remind that the spatial distribution of subrelativistic
 protons is described by the quasi-stationary diffusion equation
 \begin{equation}\label{pr_state}
\frac{\partial}{\partial E}\left( b(E) N\right)- \nabla D\nabla N
= Q(E,{\bf r})\,,
\end{equation}
where $N$ is the density of cosmic rays, $D$ is the spatial
diffusion coefficient, $dE/dt\equiv b(E)$ is the rate of energy
losses which for subrelativistic protons is determined by the
Coulomb collisions and has the form
\begin{equation}
\left(\frac{dE}{dt}\right)_{\rm i}=-\frac{ 4\pi ne^4\ln\Lambda_1}{
m\mathrm{v}_{\rm p}}\simeq \frac{k}{\sqrt{E}}\,,
\end{equation}
where $\mathrm{v}_p$ is the proton velocity, $\ln\Lambda_1$ is the
Coulomb logarithm, and $n$ is the plasma density, which  is a
function of the radius $r$ in general case. The source function is
taken in the form
\begin{equation}
Q(E)=Q\delta(E-E_{\rm esc})\delta({\bf r})\,.
\end{equation}
where $Q\simeq3\times 10^{45}$ protons s$^{-1}$ and $E_{\rm
esc}\simeq 100$ MeV is the injection energy of protons.

For $n=n_0=const$,  the spherically symmetric solution is
\begin{equation}
N(r,E)={{Q}\over{\mid b(E)\mid}(4\pi\lambda)^{3/2}}
\exp\left[-\frac{{\bf r}^2}{4\lambda}\right] \label{N_uni}
\end{equation}
where
\begin{equation}
\lambda(E)=\int\limits^E_{\rm
E_0}{{D(E)}\over{b(E)}}dE=\frac{2D}{3k}\left(E_{\rm
esc}^{3/2}-E^{3/2}\right)
\end{equation}

In the case when the plasma density is spatially variable as
\begin{equation}
n(r)=n_0\left(\frac{r_0}{r}\right)^2,
\end{equation}
 {where $r_0$ is a scale parameter,} the analytical solution for $N(r,E)$ is
\begin{equation}
N(r,E)={{Q}\over{\mid
b(E)\mid}r_0^2(4\pi\lambda)^{1/2}}\sqrt{\frac{r_0}{r}}\exp\left[-\frac{\lambda(E)}{r_0^2}\right]
\exp\left[-\frac{{\ln^2\left(r/r_0\right)r_0^2}}{4\lambda(E)}\right]
\label{n_nuni}
\end{equation}
 Here
\begin{equation}
\lambda=\frac{1}{4}\int\limits^E_{\rm E_0}{D(E)\over{b(E)}}dE
\end{equation}

\begin{figure}[h]
\begin{center}
\FigureFile(110mm,80mm){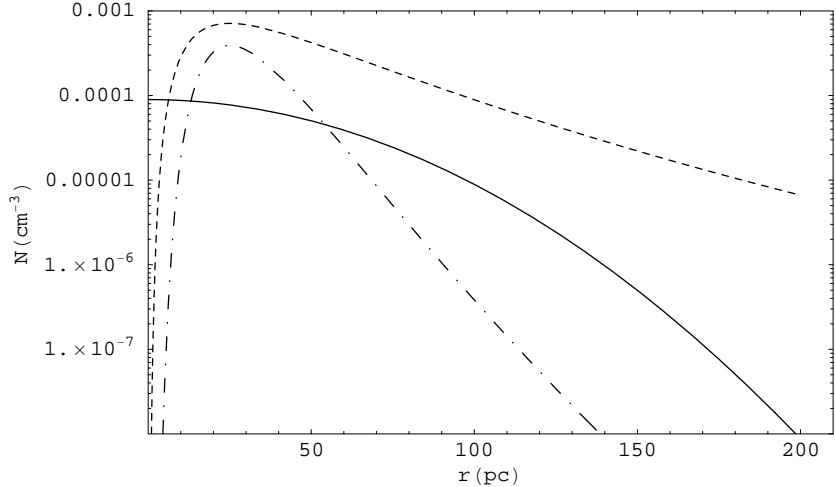}
\end{center}
\caption{The spatial distribution of 70 MeV protons in the GC for
the case of uniform distribution (solid line) and that of
non-uniform distribution equation (\ref{n_nuni}) for $n_0=10$ cm$^{-3}$
(dashed-dotted line)  and $n_0=3$ cm$^{-3}$ (dashed line).}
\label{model_distr}
\end{figure}

In figure \ref{model_distr} we show the  difference between
uniform for $n_0=0.2$
 cm$^{-3}$(solid line) and non-uniform cases
  for the parameters $n_0=10$ cm$^{-3}$ (dashed-dotted line) and $n_0=3$
 cm$^{-3}$ (dashed line).  {The scale parameter is taken to be $r_0=25$~pc.}
 One can see
 that variations of the gas density change significantly the
 proton distribution.

\section{How to compare the model calculation and the observed HXD data?}
In section 5 and 6, we will compare spectra expected from the
present model with the HXD spectrum observed around the GC
(\cite{yuasa}). Before describing the comparison result, we
present methods and assumptions which were used in the
calculation.

The intensity  {$I$} of inverse bremsstrahlung  emission of
protons
 in any direction ${\bf s}$ is calculated from
\begin{equation}
I(E_{\rm x},{\bf{ s}})=\frac{1}{4\pi}\int\limits_{ \bf
s}n(r)ds\int\limits_{E}N(E,r)v\frac{d\sigma_{\rm ib}}{dE_{\rm
x}}dE \label{is}
\end{equation}
where the integration is along the line of sight ${\bf s}$ and
$E_{\rm x}$ is the energy of photons. Here $N(E,r)$ is the density
of subrelativistic protons with the energy $E$ at the radial
distance $r$ from the GC, $v$ is the velocity of protons with the
energy $E$, $d\sigma_{\rm ib}/dE_{\rm x}$ is the cross-section of
inverse bremsstrahlung radiation, and $n(r)$ is the plasma density
distribution in the GC.

To compare such results of our model calculations, which are given
in {\it photons cm$^{-2}$keV$^{-1}$s$^{-1}$sr$^{-1}$} with the {
Suzaku} data of \citet{yuasa} given in {\it counts s$^{-1}$}, we
should convolve the HXD energy and angular responses (see
\cite{mitsu07,takaha,kokub}) to the model intensity. The procedure
is as follows:
\begin{equation}
f_{\rm 14-40}=\int_{\ell, b} d\Omega \int_{\rm 14 keV}^{\rm 40
keV} dE A(\ell, b) S(E) I(E, \ell, b)\,,
\end{equation}
where $f$ is in ({\it counts} {\it s}$^{-1}$) denotes the expected
count rate in the direction determined by the galactic coordinates
$({\it \ell,b})$, $\Omega$ is in ({\it sr}), $I(E, \ell, b)$ in
({\it {ph cm$^{-2}$keV$^{-1}$s$^{-1}$sr$^{-1}$}}) is  the
continuum intensity in the direction $(\ell, b)$, $S(E)$ and
$A({\it \ell,b})$ represent the effective area and angular
transmission of the HXD, respectively.

The HXD data can still include a contaminating flux from a number
of X-ray point sources besides the diffuse emission as noted in
Dogiel et al. (2009b).  {The flux level of the contamination is
calculated to be on the order of 10\% by integrating the known
$\rm{Log} N-\rm{Log} S$ curve for X-ray point sources around the
GC obtained with Chandra \citep{muno09} over the luminosity of of
$2\times10^{32}-1\times10^{34}$~erg s$^{-1}$ in which range has
been actually measured so far (in the "field" region of
\cite{muno09}). Although the contribution can be larger if the
$\rm{Log} N-\rm{Log} S$ curve is measured further smaller
luminosities, in the present comparison with the model
calculation, we simply ignore their contribution. The effect of
this neglection is discussed later in section 6.}

 \section{Spatial Variations of the Gas Density in the GC}
 The spatial distribution of the hot plasma in the GC can be
 derived from the distribution of the 6.7 keV iron line which
 traces the hot plasma.

The origin of the K-$\alpha$ iron line from the hot plasma of the
GC is pure thermal (see \cite{dog_pasj}). Then for the surface
brightness distribution of  the 6.7 keV emission line in any
direction ${\bf s}$ to the GC we have
\begin{equation}
I_{\rm 6.7}({\bf s})\propto \int\limits_{\bf s}n^2(r)ds
\label{i6.7}
\end{equation}

\citet{maeda8} obtained a surface brightness distribution of  the
6.7 keV emission line, $I_{\rm 6.7}$, from the ASCA data as a
function of the angle from the Galactic center, $\theta$,
\begin{eqnarray}
&&I_{\rm 6.7}(\ell,b) = I_1 \exp(-\frac{|\theta|}{\omega_1}) +
I_2\exp(-
 \frac{|\theta|}{\omega_2})\\
&& \cos \theta = \cos \ell \cos b \nonumber\end{eqnarray} where
$\theta $ is the angle from $(l,b)=(0,0)$, $I_1$=19.7 and
$I_2$=1.6 in {\it photons cm$^{-2}$s$^{-1}$sr$^{-1}$} units, and
$\omega_1=0.42$ and $\omega_2=15$ in degrees.

\begin{figure}[h]
\begin{center}
\FigureFile(110mm,80mm){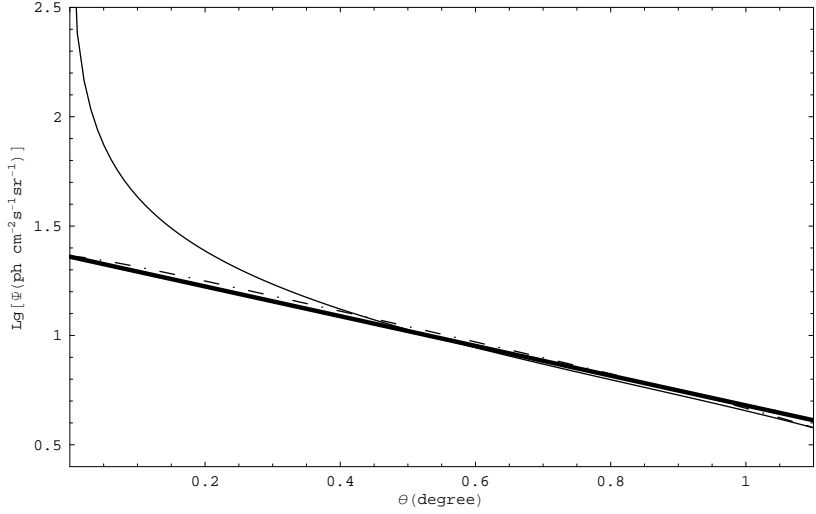}
\end{center}
\caption{The latitude distribution of 6.7 keV line as  from the
ASCA data by\citet{maeda8} is shown by the solid line. Model
spatial distribution of the surface brightness profile of  the 6.7
keV emission line expected from equation (\ref{6.7_mod}) for plasma
distributions: $r_0=25$ pc and $\alpha=0.7$ (dashed-dotted line),
and  uniform plasma distribution (thin solid line).} \label{6.7}
\end{figure}

This distribution is shown in figure \ref{6.7} by the thick solid
line.

 From the 6.7 line distribution of \citet{maeda8} we try to
derive the plasma density distribution in the GC. Then we will use
this density distribution for  calculations of hard X-ray emission
from the GC. We represent the hot plasma distribution by
analytical functions as
\begin{equation}
n(r)=n_0\exp\left[-\frac{1}{2}\left(\frac{r}{r_0}\right)^\alpha\right]\,,
\label{nr}
\end{equation}
and our goal is to find suitable parameters $r_0$ and $\alpha$
which give good correspondence of the observed ASCA data and that
obtained with the density distribution (\ref{nr}).

\begin{figure}[h]
\begin{center}
\FigureFile(110mm,80mm){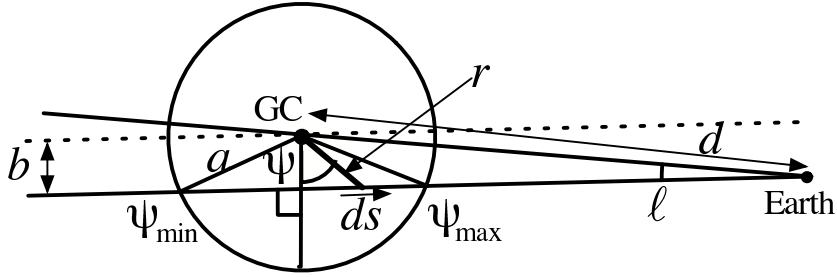}
\end{center}
\caption{The schematic view of the GC from Earth. The arrow line
shows the distance between Earth and the GC, the solid line is the
line of view, the solid circle  shows the region of hot plasma
around the GC; the two radius-vectors mark positions of $\psi_{\rm
max}$ and $\psi_{\rm min}=\pi-\psi_{\rm max}$. }\label{fig_1}
\end{figure}
In order to derive the model spatial variations of the 6.7 keV
line emission along the Galactic plane we should integrate along
the direction ${\bf s}$ shown in figure \ref{fig_1}. The angle
between the direction to the GC and the line of view
 is the Galactic longitude ${\ell}$. The circle line
around the Galactic center defines a sphere filled with a hot
plasma whose radius is supposed to be about $a=200$ pc in
accordance with the X-ray observations (see, e.g. \cite{koya1}).
The radius-vector $r$ from the GC along the line of sight  as a
function of the Galactic longitude $\ell$ and the angle $\psi$
between the line coinciding Earth an the GC (arrow line)  is
described as
\begin{equation}
r({\ell},\psi)=d\cdot\frac{\sin{\ell}}{\cos\psi}
\end{equation}
Then  with the density distribution (\ref{nr})  the integration
along the line of sight in the direction {\bf s} is
\begin{equation}
I_{\rm 6.7}\propto\int\limits_{\pi-\psi_{\rm max}}^{\psi_{\rm
max}}n^2(r(\ell,\psi))\frac{ds}{d\psi}d\psi \label{6.7_mod}
\end{equation}
where
\begin{equation}
\frac{ds}{d\psi}=-a\cdot\frac{\cos{\psi_{\rm max}}}{\cos^2\psi}
\end{equation}
and the angles $\pm\psi_{\rm max}$ shown by the two radius-vectors
in figure \ref{fig_1} is
\begin{equation}
\psi_{\rm max}=\arccos\left[\frac{d}{a}\sin{\ell}\right]
\end{equation}
Here $d=8$ kpc is the distance between Earth and the GC which
equals  8 kpc \citep{reid}. From equation (\ref{6.7_mod}) and the ASCA
6.7 keV spatial variations obtained by \citet{maeda8}  we derived
parameters $r_0$ and $\alpha$, that gives $r_0=25 - 75$ pc and
$\alpha=0.7 - 1.7$. The ASCA spatial variations of 6.7 keV line
 are shown in figure \ref{6.7} by the thick solid line. As an
example we show also in this figure the model spatial variations
of the 6.7 keV line intensity derived for $r_0=25$ pc and
$\alpha=0.7$ (dashed-dotted line). The coincidence between
calculations and data are reasonably good.  For comparison we
showed also in this figure the expected 6.7 keV line spatial
variation for the case of uniform plasma distribution (thin solid
line). One can see that the ASCA data can hardly be described by a
model with the uniform plasma distribution.

\begin{figure}[h]
\begin{center}
\FigureFile(80mm,80mm){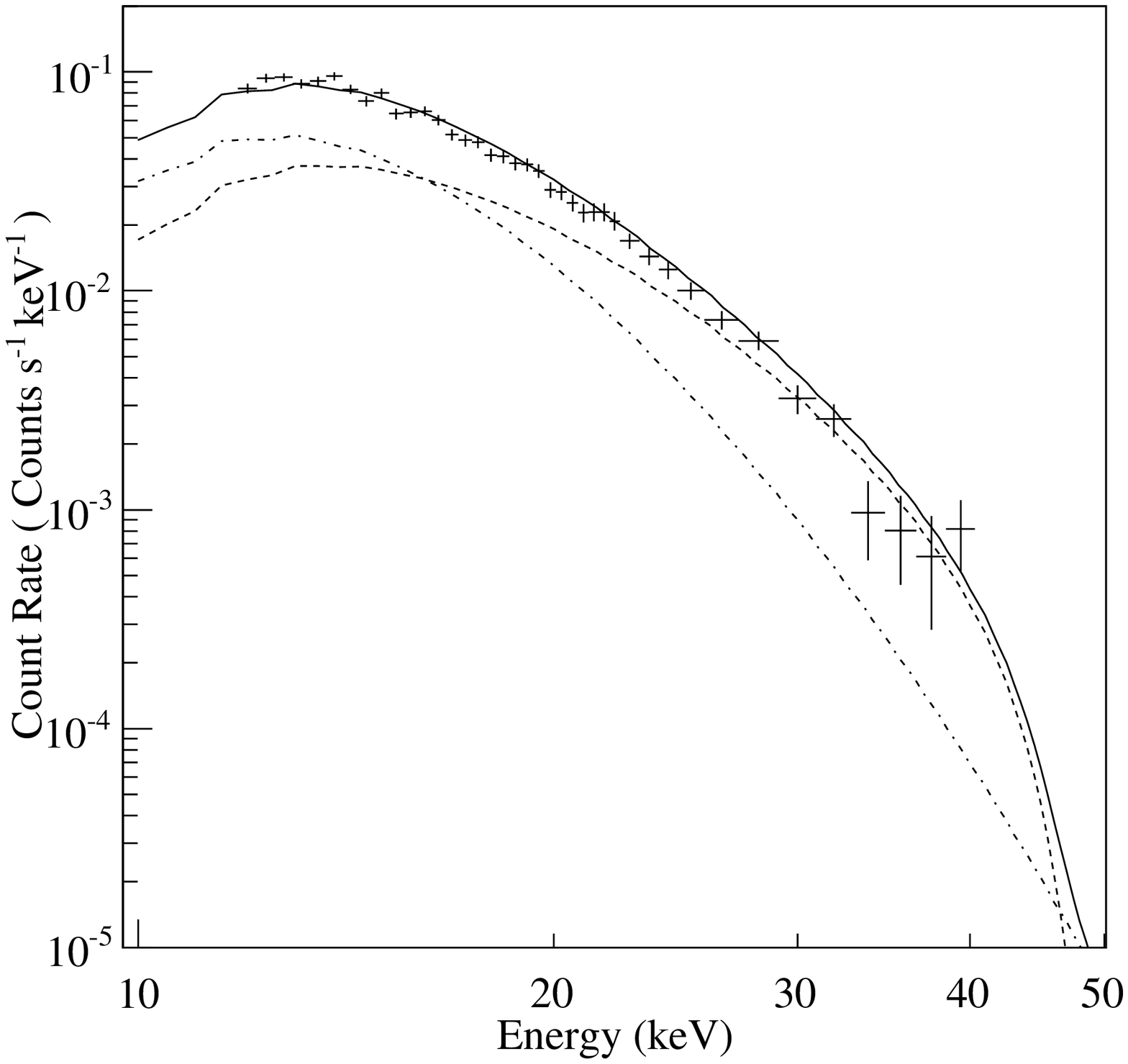}
\FigureFile(80mm,80mm){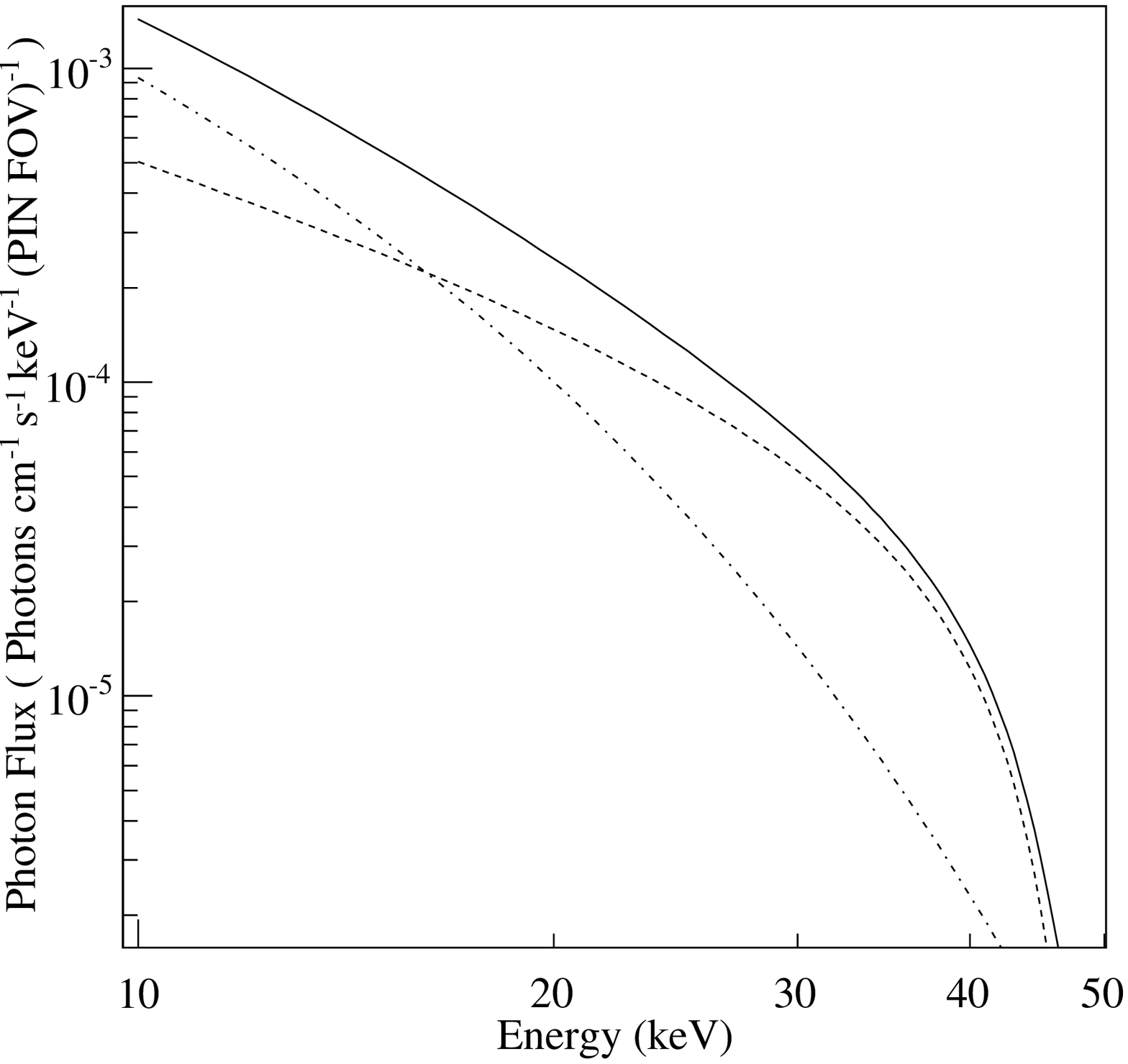}
\end{center}
\caption{
(a)The $14-40$~keV X-ray spectrum observed
by HXD/PIN around the GC (taken from Yuasa et al. 2008).
The model spectra for 6.5~keV thermal emission,
inverse bremsstrahlung, and sum of them are also shown
in dashed, dashed-dotted, and solid line. The model spectra
were calculated using such parameters as $r_0=50$~pc,
$\alpha=1.1$, $D=1\times10^{26}$~cm$^2$/s, and $n_0=0.13$~cm$^{-3}$.
(b)The same model spectra as (a) are shown without
convolving the HXD/PIN energy response but with integration
over the PIN FOV ($\timeform{0.55D}\times\timeform{0.55D}$, effectively).
}\label{XR-sp}
\end{figure}

 The parameter $n_0$ of equation (\ref{nr}) can be roughly
estimated by comparing the model emission (6.5~keV thermal +
non-thermal) with the observed spectrum of the hard X-ray emission
around the GC (\cite{yuasa}). To perform such a comparison, we
convolved the model spectrum with the HXD response as described in
section 4, and then, overlaid it over the HXD spectrum by
adjusting the model normalization factor, $n_0$ in equation
(\ref{nr}), so that the model reproduces the HXD fluxes. Figure
\ref{XR-sp}(a) shows a result of the convolution, and the model
well reproduce the observed spectrum in the $14-40$~keV band.
Although we only present the result for one parameter set
explained in the caption of figure 4, the model spectral shape
does not strongly depend on the parameters, and within possible
parameter ranges, $n_0$ was calculated to be $\sim0.1-0.3$
cm$^{-3}$.

As one can see from these figures the contributions of thermal and
non-thermal emission into the total flux from the GC are almost
equal to each other in this energy range (their ratio is 2:3).

\section{Diffusion Coefficient of Cosmic Rays in the Galactic
Center}

In order to calculate the spatial distribution of subrelativistic
protons in the GC we should take into account energy losses in
regions of hot plasma and in the regions of neutral gas
surrounding the GC whose average density was taken to be $n_H=1$
cm$^{-3}$.

As follows from the analysis of \citet{yuasa} the spatial
distribution of the intensities of thermal and non-thermal
emission in the GC are similar to each other. Since the first one
is proportional $I_{\rm th}\propto\int\limits_r n^2(r)ds$ and the
second one is proportional to $I_{\rm nth}\propto\int\limits_r
n(r)N_p(r)ds$, where $N_p(r)$ is the spatial distribution of
subrelativistic protons, it means that the spatial distribution of
subrelativistic protons does not differ strongly from that of
plasma.

\begin{figure}[h]
\begin{center}
\FigureFile(110mm,80mm){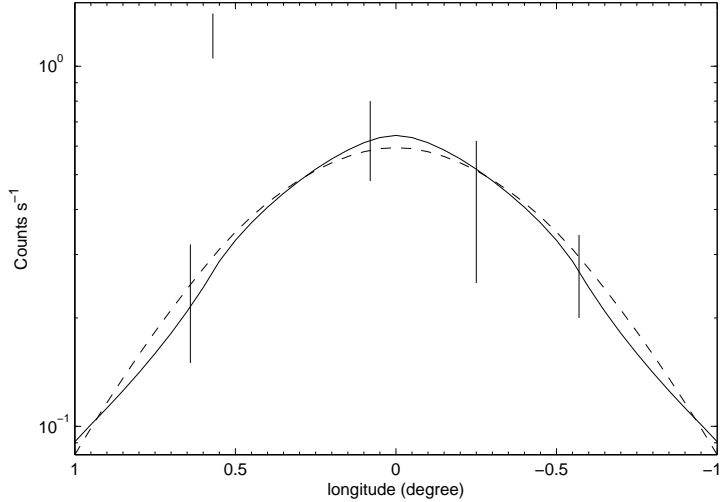}
\end{center}
\caption{The spatial distribution of inverse bremsstrahlung flux
in the energy range $12-40$~keV averaged over the HXD field of
view with the {  Suzaku} data. Solid line: $D=3\times
10^{25}$cm$^2$s$^{-1}$, $Q=10^{45}$protons s$^{-1}$.  Dashed line:
$D=3\times 10^{26}$cm$^2$s$^{-1}$, $Q=2.3\times 10^{45}$protons
$s^{-1}$. The {  Suzaku} data were taken from
\citet{yuasa}}\label{f3}
\end{figure}

The expected spatial distributions of the $14-40$~keV X-ray
emission in the GC derived from equation (\ref{is}) for $D=3\times
10^{25}$cm$^2$s$^{-1}$ (solid line) and for $D=3\times
10^{26}$cm$^2$s$^{-1}$ (dashed line) and the {  Suzaku} data
obtained by \citet{yuasa}   are shown in figure \ref{f3}
\footnote{In figure \ref{f3}, the data point at
$l=\timeform{0.55D}$ is largely discrepant from the model, and
even from the other data points. As noted in \citet{yuasa}, this
deviation could be due to underestimation of contaminating signals
from known bright point source inside the HXD/PIN FOV. Therefore,
this point should be ignored in the present comparison between
observed data and the model calculation.}. Our calculations show
that spatial distribution of the non-thermal $14-40$~keV X-ray
emission does not differ significantly from each other. However,
the necessary proton production rate $Q$ is an essential function
of the diffusion coefficient. Thus, for $D=3\times
10^{25}$cm$^2$s$^{-1}$ the value of $Q$ needed  to reproduce the {
Suzaku} data  equals $ 10^{45}$protons s$^{-1}$
 while for $D=3\times
10^{26}$cm$^2$s$^{-1}$ the production rate is necessary to be
$Q=2.3\times 10^{45}$protons s$^{-1}$. We remind that the average
rate of proton production by accretion on the black hole is no
more than $Q=3\times 10^{45}$protons s$^{-1}$. Since the value of Q
is restricted,  then permitted values of  $D$ are confined within
a relatively narrow range  around
 $D\simeq 10^{26}$ cm$^2$ s$^{-1}$.

However, these estimates were obtained in assumption that almost
100\% of the $14-40$~keV X-ray flux from the GC is generated by
subrelativistic protons. As follows from, e.g., \citet{warw,rev09}
a significant part of the GC hard X-ray flux may be due to faint
point sources. Then a smaller production rate of protons in the GC
is required, and, hence, higher values of the diffusion
coefficient cannot be excluded.

 \section{Discussion and  Conclusion}

 Analysis of X-ray data \citep{koya09} demonstrated that continuum non-thermal emission in the
 GC consisted of two components one of which was proportional to the
 intensity of 6.7 keV iron line which traced distribution of hot 6.5 keV
 plasma, and the other one was proportional to the intensity of 6.4
 keV plasma which traced distribution of cold molecular hydrogen.
 If this non-thermal emission is due to inverse bremsstrahlung
 emission of subrelativistic protons, then their interactions with
 hot and cold fractions of the interstellar medium are equally
 important. Our estimation show that about 30\% of the total
 non-thermal flux from the GC in the range $14-40$~keV is generated in
 regions of cold gas while the rest is produced by proton
 interaction with hot plasma. The reason is that only a small
 fraction of molecular gas is involved into processes of
 bremsstrahlung emission because of short mean free path of protons
 in molecular clouds.

 From the X-ray data obtained with
ASCA  \citep{maeda8} it was concluded that the plasma density
distribution in the GC is strongly non-uniform, e.g. the model
distribution of 6.7~keV line in the GC obtained with the plasma
profile
\begin{equation}
n(r)=0.2\times
\exp\left[-\left(\frac{r}{50(\mbox{pc})}\right)^{1.1}\right](\mbox{cm$^{-3}$})
\end{equation}
is very close to that from ASCA observation. This effect of
density variations should be included into analysis of protons
propagation in the GC.

 From the total spectrum of X-ray emission in the range $14-40$~keV
 from the $0.55^\circ$-diameter central region and the spatial
 variation of this emission within $1.5^\circ$-diameter central
 region as observed by {  Suzaku} and {  ASCA} we derived  the characteristic diffusion
 coefficient of subrelativistic protons. It was shown that  the diffusion coefficient of
 subrelativistic particles is about the value of $3\cdot 10^{26}$
 cm$^2$s$^{-1}$. For  higher values of the diffusion
 coefficient a higher production rate of subrelativistic protons is
 necessary which cannot be provided by accretion.  However, this estimation of
 $D$ one should consider as a lower limit. The point is that we
 assumed here that all hard X-ray emission in the $14-40$~keV band is produced by
subrelativistic protons in the hot fraction of the interstellar
gas. However as stated above  a
 part of this emission may be produced in molecular clouds.
 Besides, we do not know exactly which part of this emission is
due to  unresolved point sources (see discussion in
\cite{dog_pasj}). Therefore, we cannot exclude that the diffusion
coefficient in the GC is close to its average value in the Galaxy,
i.e. $D\sim 10^{27}$ cm$^2$ s$^{-1}$. Thus, we determine the range
of values of the diffusion coefficient in the GC as $10^{26} -
10^{27}$cm$^2$s$^{-1}$.

\vspace{5 mm}  The authors are grateful to the  referee, Prof.
Fumiaki Nagase, for his comments and corrections.

 VAD and DOC were partly supported by the RFBR
grant 08-02-00170-a, the NSC-RFBR Joint Research Project No
95WFA0700088 and by the grant of a President of the Russian
Federation "Scientific School of Academician V.L.Ginzburg". KSC is
supported by a RGC grant of Hong Kong Government under HKU
7014/07P. A.~Bamba is supported by JSPS Research Fellowship for
Young Scientists (19-1804). CMK is supported in part by National
Science Council, Taiwan under the grant NSC-96-2112-M-008-014-MY3.




\begin{thebibliography}{}

\bibitem[Ayal et al.(2000)]{ayal}
Ayal, S., Livio, M., \&  Piran, T. 2000, ApJ, 545, 772
\bibitem[Cheng et al.(2006)]{cheng1}
Cheng, K. S., Chernyshov, D. O., \& Dogiel, V. A. 2006, \apj, 645,
1138.

\bibitem[Cheng et al.(2007)]{cheng2}
Cheng, K. S., Chernyshov, D. O.,  \& Dogiel, V. A. 2007, \aap,
473, 351.

\bibitem[Dogiel et al.(2008)]{dog08}
Dogiel, V.A., Cheng, K.-S., Chernyshov, D.O. et al. 2008, New
Astronomy Reviews, 52, 460
\bibitem[Dogiel et al.(2009a)]{dog09}
Dogiel, V. A., Tatischeff,V.,  Cheng, K.-S., et al. 2009a, A\&A,
submitted
\bibitem[Dogiel et al.(2009b)]{dog_pasj}
Dogiel, V.,  Chernyshov D., Yuasa, T. et al.   2009b, PASJ,
submitted
 \bibitem[Dogiel et al.(2009c)]{dog_p3}
Dogiel, V. A.,  Cheng, K-S., Chernyshov, D.O. et al.
  2009c, PASJ, to be published in PASJ, 61, No.5
  \bibitem[Donley et al.(2002)]{don}
Donley, J. L., Brandt, W. N., Eracleous, M., \& Boller, Th. 2002,
ApJ, 124, 1308.
\bibitem[Ginzburg(1989)]{ginz}
Ginzburg, V.L. 1989, {\it Applications of Electrodynamics in
Theoretical Physics and Astrophysics}, Gordon and Brech Science
Publication.
\bibitem[Kokubun et al.(2007)]{kokub}
Kokubun, M., Makishima, K., Takahashi, T. et al. 2007, PASJ, 59,
53
\bibitem[Koyama et al.(1996)]{koya1}
 Koyama, K., Maeda, Y., Sonobe, T. et al.
1996, \pasj, 48, 249
\bibitem[Koyama et al.(2007a)]{koya3}
Koyama, K., Inui, T., Hyodo, Y. et al. 2007a, PASJ, 59, 221
\bibitem[Koyama et al.(2007b)]{koya2}
Koyama, K., Hyodo, Y., Inui, T. et al. 2007b, PASJ, 59, S245

\bibitem[Koyama et al.(2009)]{koya09}
Koyama, K., Takikawa, Y., Hyodo, Y. et al. 2009, PASJ, 61, S255
\bibitem[Maeda(1998)]{maeda8}
Maeda, Y. 1998, Ph.D.~Thesis

\bibitem[Mezger et al. (1996)]{mezger}
Mezger, P. G., Duschl, W. J.,  \& Zylka, R. 1996,
Astro.Astrophys.Rev., 7, 289

\bibitem[Mitsuda et al.(2007)]{mitsu07}
Mitsuda, K., Bautz, M., Inoue, H. et al. 2007, PASJ, 59, S1


\bibitem[Muno et al.(2009)]{muno09}
Muno, M.P., Bauer, F.E., Baganoff, F.K. et al. 2009, ApJS, 181,
110


\bibitem[Reid(1993)]{reid}
Reid, M. J. 1993, ARA\&A, 31, 345
\bibitem[Revnivtsev et al.(2009)]{rev09}
Revnivtsev, M., Sazonov, S., Churazov, E. et al. 2009 Nature, 458,
1142
\bibitem[Syer \& Ulmer(1999)]{syer}
 Syer, D., \& Ulmer, A. 1999 MNRAS, 306, 35
\bibitem[Takahashi et al. (2007)]{takaha}
Takahashi, T., Abe, K., Endo, M. et al. 2007, PASJ, 59, 35

\bibitem[Warwick et al.(2006)]{warw}
Warwick, R., Sakano, M., \& Decourchelle, A. 2006, Journal of
Physics Conference Series, 54, 103
\bibitem[Yuasa et al.(2008)]{yuasa}
Yuasa, T., Tamura, K., Nakazawa, K. et al. 2008, PASJ, 60, S207
\end{thebibliography}
\end{document}